\documentclass[submission,copyright,creativecommons]{eptcs}

\providecommand{\event}{ThEdu 2024} 

\usepackage{underscore}         
\usepackage[T1]{fontenc}        

\usepackage[utf8]{inputenc}
\usepackage{amssymb, amsmath, amsthm, mathtools}
\usepackage{amsfonts, bbold, dsfont}
\allowdisplaybreaks

\usepackage{listings, lstcoq, parcolumns}
\newcommand{\code}[1]{\mbox{\texttt{#1}}}

\newcommand{\AuthorsList}{%
  S. Boldo, F. Clément, D. Hamelin, M. Mayero \& P. Rousselin}
\newcommand{\copyrightholders}{%
  Boldo, Clément, Hamelin, Mayero \& Rousselin}

\newcommand{\LMFa}{%
  \begin{tabular}{c}
    Université Paris-Saclay, Inria,\\
    CNRS, ENS Paris-Saclay, LMF\\
    91190 Gif-sur-Yvette, France.
  \end{tabular}}
\newcommand{\SERENAa}{%
  \begin{tabular}{c}
    Inria\\
    48 rue Barrault, 75013 Paris.\\
    CERMICS, École des Ponts\\
    77455 Marne-la-Vallée, France.
  \end{tabular}}
\newcommand{\PXIIIa}[1]{%
Université Sorbonne Paris Nord, CNRS,\\
{#1},\\
F-93430 Villetaneuse, France.}
\newcommand{\LIPNa}{%
  \begin{tabular}{c}
    \PXIIIa{Laboratoire d'Informatique de Paris Nord, LIPN}
  \end{tabular}}
\newcommand{\LAGAa}{%
  \begin{tabular}{c}
    \PXIIIa{Laboratoire Analyse, Géométrie et Applications, LAGA}
  \end{tabular}}

\newcommand{\Funding}{%
  This work was supported by the Inria Challenge LiberAbaci.\protect\\
  This work was also partly supported by the European Research Council (ERC)
  under the European Union's Horizon 2020 Research and Innovation Programme –
  Grant Agreement n$^\circ$810367.
}

\newcommand{\Title}{Teaching Divisibility and Binomials with {\Coq}}

\newcommand{\Abstract}{%
  The goal of this contribution is to provide worksheets in {\Coq} for students
  to learn about divisibility and binomials.
  These basic topics are a good case study as they are widely taught in the
  early academic years (or before in France).
  We present here our technical and pedagogical choices, the numerous
  exercises we developed and a small experiment we conducted on two students.
  As expected, it required additional {\Coq} material such as other lemmas and
  dedicated tactics.
  The worksheets are freely available and flexible in several ways.
}

\newcommand{\myfootref}[1]{\textsuperscript{\ref{#1}}}

\newcommand{\apriori}{a priori}

\newcommand{\eg}{e.g.}
\newcommand{\ie}{i.e.}

\newcommand{\etal}{\emph{et al.}}

\newcommand{\Euclidean}{Euclidean}

\newcommand{\soft}[1]{\textsf{#1}}
\newcommand{\module}[1]{\soft{#1}} 
\newcommand{\Coq}{\soft{Coq}}

  \newcommand{\Reals}{\soft{Reals}}
  \newcommand{\CoqIde}{\soft{CoqIde}}
  \newcommand{\jsCoq}{\soft{jsCoq}}

  \newcommand{\Coquelicot}{\soft{Coquelicot}}
  \newcommand{\MathComp}{\soft{math-comp}}

  \newcommand{\Bertrand}{\soft{Bertrand}}
  \newcommand{\coqprime}{\soft{coqprime}}
  
  \newcommand{\FSets}{\soft{FSets}}
  \newcommand{\RSA}{\soft{RSA}}

  \newcommand{\NatBinomial}{\module{Nat\_Binomial}}
  \newcommand{\NatCompl}{\module{Nat\_Compl}}
  \newcommand{\NatFactorial}{\module{Nat\_Factorial}}
  \newcommand{\NatSum}{\module{Nat\_Sum}}
  \newcommand{\WSHelper}{\module{WS\_Helper}}
  \newcommand{\WSBinomial}{\module{WS\_Binomial}}

\renewcommand{\leq}{\leqslant}
\newcommand{\N}{\mathbb{N}}
\newcommand{\Z}{\mathbb{Z}}
\newcommand{\R}{\mathbb{R}}

\newcommand{\eqdef}{\stackrel{\mathrm{def.}}{=}}
\newcommand{\EQDEF}{\,\eqdef\,}
\newcommand{\OR}{\lor}
\newcommand{\IMPLIES}{\quad\Longrightarrow\quad}

\newcommand\french\textsl

\usepackage[normalem]{ulem}

\usepackage[color=orange]{todonotes}

\definecolor{darkred}{rgb}{0.8,0.2,0.2}
\definecolor{darkgreen}{rgb}{0.2,0.8,0.2}
\definecolor{darkblue}{rgb}{0.2,0.2,0.8}
\definecolor{lightred}{RGB}{255,225,225}
\definecolor{lightgreen}{RGB}{200,255,200}
\definecolor{lightblue}{RGB}{225,225,255}

\title{\Title\thanks{\Funding}}
\def\titlerunning{\Title}

\author{%
  Sylvie Boldo
    \institute{\LMFa}
    \email{sylvie.boldo@inria.fr}
  \and François Clément
    \institute{\SERENAa}
    \email{francois.clement@inria.fr}
  \and David Hamelin
    \institute{\LMFa}
    \email{david.hamelin@inria.fr}
  \and Micaela Mayero
    \institute{\LIPNa}
    \institute{\LMFa}
    \email{mayero@lipn.univ-paris13.fr}
  \and Pierre Rousselin
    \institute{\LAGAa}
    \institute{\SERENAa}
    \email{rousselin@univ-paris13.fr}
}
\def\authorrunning{\AuthorsList}

\begin{document}

\maketitle

\begin{abstract}
\Abstract
\end{abstract}

\section{Introduction}
\label{sec:intro}

This work was done in the context of the LiberAbaci project%
\footnote{\url{https://liberabaci.gitlabpages.inria.fr/}} aimed at improving
the accessibility of the {\Coq} proof system for mathematics students in the
early academic years.

A difficulty in France is that there are many different curricula for the first
academic years, so it was difficult to choose which topic to cover.  A solution
was to focus on just before higher education.  At 17--18 years old, French
students are usually at the last year of high school (\french{lycée}), this
year being called \french{terminale} is ending by an exam called
\french{baccalauréat}.  With the high school reform from 2020 in France,
students may choose mathematics as major (\french{spécialité}), possibly with
additional mathematics during the last year.  We have focused on these national
curricula, more precisely on integers: divisibility and binomials.  It is not
the first time {\Coq} tackles mathematics from \french{baccalauréat}: in 2013,
the real analysis exercise was done in {\Coq} at the same time as students
\cite{Lelay15} as a test of the {\Coquelicot} library \cite{BLM15}.

As for the exercises, they follow the spirit from Software
Foundations~\cite{pierce2010software} that was also used at Sorbonne Paris-Nord
University~\cite{smf} of a {\Coq} file with holes.  We
provide exercises with comments both in French and English (as the language
should not be the main difficulty for the students).  Depending on the
difficulty of the exercise, we may provide hints or explanations.

We provide several levels of leeway to the teacher.  First, the teacher may
choose among the exercise (and their order) from the worksheets (files named
\code{WS\_*.v}).  Second, some lemmas of the core library can be seen as
exercises, such as the committee-chair identity about binomials in
Section~\ref{sec:binom-ct}.
Third, the level of automation is also up to
the teacher.  We provide all the exercise proofs without automation such as
\coqe{lia} but it may be allowed for more advanced proofs in order to shorten
them.  Last but not least, the teacher chooses the environment to be used by
the students, both the IDE (possibly {\jsCoq}) and the required
libraries (with a limited number of lemmas, with or without automation and so
on).

Our goal is that formal proofs be as near as possible to mathematical proofs.
{\Coq} is here to prevent the student from making errors, forgetting a
sub-case, and so on, as much as the student needs, and without getting on the
teacher's nerves.  We want the statements and proofs to be as near as possible
to the mathematical reasonings and pedagogy.  Even if we want to increase the
abstraction level of students, we chose not to rely on too much abstraction in
these exercises.  A first reason is that it is probably too early in the
academic years and a second reason is that error messages would be impossible
to understand.  This is why we rely on plain Peano naturals, and not on an
abstract commutative monoid (where we could have defined sums for instance).
We therefore only depend upon the {\Coq} standard library, and not on
{\MathComp} or {\Coquelicot} (for sums for instance) that abstract too much for
our purpose.  See also Section~\ref{sec:binom-soa}.

Divisibility and binomials are also commonly studied in the first academic year
by students in mathematics and the contributions of this article range from
\french{terminale} to the early academic years.  Section~\ref{sec:div} is about
divisibility and Section~\ref{sec:binom} about
binomials. Section~\ref{sec:exp} describes a small experiment on two
students in June 2024 that has led to a several improvements of the
worksheets.

All the {\Coq} material (and a useful script explained later) is freely
available at the following address:\\
\centerline{\url{https://liberabaci.gitlabpages.inria.fr/contenus/ThEdu24/ThEdu24.tar.gz}}\\
that compiles with {\Coq} versions 8.16 to 8.20.  The worksheets are also
available (with their proof for the demo) to be tested
in a browser thanks to {\jsCoq}~\cite{jscoq} at:\\
\centerline{\url{https://liberabaci.gitlabpages.inria.fr/contenus/ThEdu24/div.html}}\\
\centerline{\url{https://liberabaci.gitlabpages.inria.fr/contenus/ThEdu24/binom.html}.}

\section{Divisibility}
\label{sec:div}

The divisibility relation on integers and basic arithmetic in general are often
taught in high school and at the beginning of mathematically oriented higher
education.  Without neglecting historical
(arithmetic was, along with geometry, one of the main mathematical sciences
in the Greek antiquity)
and practical (\eg{} in cryptography) aspects, there are also important
pedagogical reasons to teach arithmetic.  Indeed, even a very basic course in
arithmetic will expose students to many proof schemes, potent and nontrivial
algorithms, and predicates of very different forms (primality, divisibility)
which require the students to design careful rigorous proofs (sadly, maybe for
the first time).  In particular, the divisibility relation has many pedagogical
benefits~: working with an existentially quantified predicate, showing the
power of symbolic calculus to prove unexpected facts and using one of many
different tools (explicit witness, remainder computation, modular arithmetics,
prime factor decomposition, and so on) to solve an exercise.  It also serves as
an example of a non-total order relation, with algebraic compatibility results.

Section~\ref{sec:div-soa} presents the state of the art and some technical
choices, notably on the use of either \coqe{N} or \coqe{Z}.  A point on
primality is given in Section~\ref{sec:prime}.  For binomials, we did not
notice any tactics missing as is the case here.  For the simplicity of proofs,
we defined several dedicated tactics and constructs in
Section~\ref{sec:div-tactics}, before describing the worksheet in
Section~\ref{sec:div-ex}.

\subsection{State of the Art and Technical Choices}
\label{sec:div-soa}

We chose definitions and basic lemmas from the {\Coq} standard
library\myfootref{fn:coq-stdlib} to back our exercises.  It has many benefits,
together with some drawbacks (which will be discussed shortly): it is widely
used, does not expose the user to many abstractions and is also used in the
series ``Software Foundations'' by Pierce \etal{}~\cite{pierce2010software}.

Another choice to make is whether we work with natural numbers or integers.
Natural numbers are easier to work with in some respects, for instance there is
no need to consider units in prime number decompositions and there is only one
possible choice for {\Euclidean} division.  However, they become impracticable
whenever subtraction might be involved, for instance in Bezout's lemma.  We
chose to stick with integers, a last argument being that the standard library
offers much more in terms of arithmetics for integer than for natural numbers.

The standard library defines (in \module{Numbers.BinNums}) the type \coqe{Z} of
integers in two steps.  First, \coqe{positive} (nonzero) natural numbers are
defined as non-empty strings of bits, starting with $1$.  Then, an integer
number in \coqe{Z} is $0$ or a \coqe{positive} number with a sign.
This representation of integers has many qualities, the first being that
equality needs not be redefined (there is no need to use a quotient as is the
case for instance with the usual mathematical construction of $\Z$ as
equivalence classes of couples of natural numbers).  The other one, maybe not
relevant for an introductory mathematics course, is that this makes
operations quite fast.

The bulk of the theory of integers in the standard library is actually
implementation independent.  (Most) lemmas and conventions which are shared
with natural numbers are proved in the \module{Numbers.NatInt} sub-theory
following an axiomatic module-based approach.  Lemmas specific to integers are
proved (still in an axiomatic way) in the \module{Numbers.Integer.Abstract}
sub-theory.  Finally, these theories are included in the concrete module
\module{ZArith.BinInt.Z}.
This axiomatic module-based approach makes it possible to factor
many lemmas which are common between natural numbers and integers and
also transfer for free all the theory, say of natural numbers, to
any concrete implementation (say Peano natural numbers or binary
natural numbers).

This module-based approach has some benefits (naming consistency, code
factoring) as well as downsides: many concrete lemmas have useless hypotheses
(it is frustrating, for instance, to have to prove $0\le n$ when $n\in\N$).
Another source of disappointment is the power operation, which is specified to
be of type \coqe{t -> t -> t}, with \coqe{t} being \eg{} Peano's natural
numbers (which is fine, then) or integers (which is certainly not
mathematically natural and adds some noise in the proofs).  In addition, some
lemmas are weaker than expected, even with this factorization effort in mind.
In practice, we have to provide a thin layer on top of the standard library to
remedy these defects.  In addition to this general theory of integers, we use
the \module{ZArith.Znumtheory} module which contains, for instance Gauss' and
Bezout's lemmas.

Finally, let us discuss {\Euclidean} division conventions on $\Z$.  This is
mostly hidden under the rug in usual mathematics courses.  Indeed the
mathematics \french{spécialité} national curriculum%
\footnote{\url{https://eduscol.education.fr/document/24574/download}.}  only
mentions {\Euclidean} division of an integer by a positive natural number.  In
programming languages there are mostly two flavors of {\Euclidean} divisions on
integers~: ``floor division'', where the quotient of $a$ by $b$ is the greatest
integer smaller or equal to the rational $a/b$ and ``trunc division'' where it
is the value of the ration $a/b$ truncated towards $0$.  This is discussed at
length in the standard library file \module{Numbers.Integer.Abstract.ZDivFloor}
which also links to the useful reference \cite{boute1992euclidean}.  We chose
the floor division provided by the standard library.  It does not have any
mathematical advantage over the others, but its notations are less surprising
and it is the sole division considered in \module{Znumtheory}.  In \Coq{},
there is no primitive notion of partial function, so one should decide the
result of the {\Euclidean} division by $0$.  The {\Coq} development team chose
wisely \coqe{\forall a : Z, a / 0 = 0} and \coqe{a mod 0 = 0}, so that the
formula \coqe{\forall a b : Z, a = b * (a / b) + a mod b} holds even when
\coqe{b = 0}.  This actually simplifies greatly many proofs.

\subsection{Computation and prime numbers}
\label{sec:prime}

One of the distinguishing traits of \Coq{} is its orientation towards
computation.  Most of the operations are actually defined by (terminating)
programs which can compute if given explicit arguments (and usually perform as
many reduction steps as they can).  For instance the \coqe{mod} operation on
\coqe{Z} actually computes the remainder of an {\Euclidean} division, using an
efficient algorithm.  Then \coqe{mod} is proved to satisfy its specification
and all the subsequent theory follows.

In contrast to computation, \coqe{Prop}-valued predicates such as the
divisibility relation in \coqe{Z} express a property which is (in general) not
obtained by computation.  In the standard library, it is expressed in a very
natural way with an existential quantifier:
\begin{lstlisting}
Definition divide n m = \exists p, m = n * p.
\end{lstlisting}
The usual notation for divisibility allows for more concise statements:
\begin{lstlisting}
Infix "|" := divide (at level 0).
\end{lstlisting}
Some lemmas serve as a bridge between the computational and the
non-computational world:
\begin{lstlisting}
Check Z.mod_div : \forall a b : Z, a mod b = 0 <-> (b | a).
\end{lstlisting}
This means that, during the course of an exercise, one can prove explicit
divisibility relations simply by computation:
\begin{lstlisting}
Lemma example_div : (31 | 62744).
Proof. apply Z.mod_div; reflexivity. Qed.
\end{lstlisting}
Here, the \coqe{reflexivity} tactic checks that \coqe{62744 mod 31} and $0$ are
\emph{convertible}, in short, in our case, that the result of the computation
of the left operand is indeed $0$.  This kind of translation between properties
and computations is the basis of the
``Small Scale Reflection'' methodology used in the
\MathComp{} library~\cite{gonthier2016small}: mathematical propositions having
two sides, one computational and one logical, with reflection lemmas (akin to
\coqe{Z.mod_div} here, but more elaborate) to toggle between those.

While doing some exercises, we noticed that the standard library lacks a
function to decide the primality of an integer.  The \coqe{prime} predicate in
\module{Znumtheory} is itself quite surprising%
\footnote{We have modified it slightly for simplification purposes.}:
\begin{lstlisting}
Definition prime (p : Z) : Prop := 1 < p /\ \forall n, 1 <== n < p -> Z.gcd n p = 1.
\end{lstlisting}
In practice, it proved to be as inconvenient as it is pedagogically debatable.
Notice the choice to consider negative numbers non-prime which contradicts more
advanced ideal based definitions of primality, but is common in introductory
mathematical courses on arithmetic.  We provide some alternative formulations,
as well as a Boolean-valued function \coqe{primeb}, which computes:
\begin{lstlisting}
Lemma prime_no_strict_divisor (p : Z) :
  prime p <-> 1 < p /\ \forall a, (a | p) -> a = -1 \/ a = 1 \/ a = p \/ a = - p.
Lemma prime_no_strict_divisor_below_sqrt (p : Z) :
  prime p <-> 1 < p /\ \forall a, 1 < a <== Z.sqrt p -> ~ (a | p).
Lemma primeb_correct (n : Z) : primeb n = true <-> prime n.
\end{lstlisting}
Our \coqe{primeb} function simply computes the remainders of the division of an
integer $p \ge 2$ by $2, 3, ..., \sqrt{p}$, where $\sqrt{p}$ is the integer
(floor) square root of $p$.  This way, it becomes possible in exercises to
prove well-known primality results by computation:
\begin{lstlisting}
Lemma prime_101 : prime 101.
Proof. apply primeb_correct; reflexivity. Qed.
\end{lstlisting}

Of course, our algorithm is quite naive and pales in comparison to
state-of-the-art methods such as what can be found, for instance, in the
{\coqprime} library~\cite{thery2007primality}.  It is simple, probably what the
students expect at this point, and can compute reasonably fast up to integers
of around 8 decimal digits, and does not require to import a large development.

\subsection{Tactics}
\label{sec:div-tactics}

After the ability to compute whether an integer is prime, another gap in the
existing tactics is the ability to easily split into various subcases.  More
precisely, a recurring theme in divisibility exercises is to study the various
possible results of a modulo.  For example, an integer has only seven
possibilities of values for its modulo~7.  This is used in the following
exercise:
\[
  \forall n > 0,\;
  \left( \exists k > 0, \exists q > 0,\; n = k^2 \wedge n = q^3 \right)
  \Rightarrow \left( n \equiv 0 \, [7] \vee n \equiv 1 \, [7] \right).
\]
It can be proven by considering the modulo~7 remainders of $k^2$ and $q^3$.
The mathematical proof is then the following table and the fact that only~0
and~1 appear in both lines.
\begin{center}
  \begin{tabular}{|l|l|l|l|l|l|l|l|}
    \hline
    $k,q$ & 0 & 1 & 2 & 3 & 4 & 5 & 6 \\
    \hline
    $k^2\mod7$ & \color{red}{0} & \color{red}{1} & 4 & 2 & 2 & 4 & 1 \\
    \hline
    $q^3\mod7$ & \color{red}{0} & \color{red}{1} & 1 & 6 & 1 & 6 & 6 \\
    \hline
  \end{tabular}
\end{center}

The {\Coq} formalization of the property is straightforward, but the proof is
tedious as we have to split many cases without any help from the tool:
\begin{lstlisting}
Lemma exercise11 (n k q : Z) (h0n : 0 <== n) (h0k : 0 <== k) (h0q : 0 <== q)
    (hk : n = k * k) (hq : n = q * q * q) : n mod 7 = 0 \/ n mod 7 = 1.
Proof. assert (n mod 7 = 0 \/ n mod 7 = 1 \/ n mod 7 = 2 \/ n mod 7 = 4) as h1.
- rewrite !hk, Z.mul_mod.
  destruct (Z.eq_dec (k mod 7) 0) as [hk0|hk0].
  + rewrite hk0. tauto. (* solves the goal for k mod 7 = 0 *)
  + destruct (Z.eq_dec (k mod 7) 1) as [hk1|hk1].
    * rewrite hk0. tauto. (* solves the goal for k mod 7 = 1 *)
(*... same for 2,3,4,5 and 6: we have k mod 7 <> 0, k mod 7 <> 1, ... k mod 7 <> 6... *)
    * pose proof (mod_pos_bound k 7); nia. (*... but k mod 7 < 7, so the goal is absurd *)
- assert ((n mod 7) = 0 \/ (n mod 7) = 1 \/ (n mod 7) = 6) as h2. (* ... same as h1 *)
lia. Qed.
\end{lstlisting}
To solve this issue, we have added the \coqe{study} tactic, that fits the table
construction by making the adequate splitting.  It considerably shortens the
proof:
\begin{lstlisting}
assert (n mod 7 = 0 \/ n mod 7 = 1 \/ n mod 7 = 2 \/ n mod 7 = 4) as h1.
- rewrite !hk, Z.mul_mod.
  study (k mod 7) between 0 and 6 as h2.
  8: pose proof (Z.mod_pos_bound k 7) as bounds; lia.
  all: rewrite <-h2; tauto.
- assert (n mod 7 = 0 \/ n mod 7 = 1 \/ n mod 7 = 6) as h2.
  + rewrite !hq, (Z.mul_mod (q * q)), (Z.mul_mod q q).
    study (q mod 7) between 0 and 6 as h2.
    8: pose proof (Z.mod_pos_bound q 7) as bounds; lia.
    all: rewrite <-h2; tauto.
  + lia.
\end{lstlisting}

Generating a large number of {\Coq} goals may seem hard to handle for students,
but it should motivate them to learn selectors such as \coqe{all:} or
\coqe{1,2:} in order to avoid repetition in the proofs.

\begin{lstlisting}[caption={An example from exercise15}]
study x between 0 and 8 as h3.
10: apply Z.divide_pos_le in h2; lia.
all: subst x.
1: now apply Z.divide_0_l in h2.
1, 2, 4, 8: now repeat constructor.
\end{lstlisting}

\paragraph{Finding all solutions}
This is not a tactic but a definition that aims to stick to mathematical
exercise statements.  They will often ask to determine the possible numbers
satisfying a given property; for instance ``\textit{Find all natural numbers
  such that n+8 is a multiple of n}''.  Such statements imply that there exists
a finite set satisfying the property: it is the student's job to construct it.
To express those problems, we added a new definition \coqe{findall}, used as
follows:
\begin{lstlisting}
Lemma exercise15 : findall (\fun n => 0 <== n /\ n | n + 8).
\end{lstlisting}
where \coqe|findall {E} (P : E -> Prop) := { l : list E & Forall P l /\ \forall (x : E), P x -> In x l }|.
In other\linebreak[4] words, the student must provide a list of all the elements satisfying
the given property.  We use a Sigma-type, which lives in \coqe{Type} rather
than \coqe{exists} which lives in \coqe{Prop}: this forces the student to
actually construct the object, rather than using a proof by contradiction.

\subsection{Worksheet}
\label{sec:div-ex}

We were surprised by the variety and number of exercises we found about
divisibility.  This has led to a large variety in the formalization and
pedagogy of the formal counterpart.  Some exercises were straightforward to
formalize; for instance, \coqe{(6 | n) <-> (3 | n) /\ (2 | n)} is stated
exactly like we would on paper, and the formal proof uses the same arguments.
Some of the exercises become straightforward thanks to additional lemmas and
the use of the tactics explained in Section~\ref{sec:div-tactics}, such as
``find all the $n$ so that $n\ |\ n+8$''.

We found out exercises become pedagogical failures when formalized.  This
is the case of computational exercises such as \coqe{17 | 35^228 + 84^501}:
{\Coq} computes the modulo without the students having to think about a way to
do it themselves.  We think one such exercise is interesting for outlining
computation, but they miss how to cleverly compute $a^b \mod c$.

Many exercises rely on the decimal representation of number, which is out of the
scope of the standard library.  We therefore developed definitions and lemmas
to state results such as the following ones: 91 divides any number written
$abcabc$; 2 divides $n$ if and only if 2 divides the last digit of $n$; 3
divides $n$ if and only of 3 divides the sum of the digits of $n$.  We chose to
define our own inductive type \coqe{Digits} rather than using a list of digits,
so that the student will not have to deal with empty digits list (for the last
digit for instance).\\
We explored the option of only allowing a canonical representation of decimal numbers by forbidding trailing zeroes. This proved to be a bad idea in practice, so we went back to a simpler representation. Another consideration was choosing whether our list of digits was little or big endian. Both approaches have strengths and weaknesses; it is easier to convert a little endian digit list into a number :

\noindent\begin{minipage}{.55\textwidth}
\begin{lstlisting}[caption=Little endian]{littleendian}
Fixpoint Z_of_Digits_alt (l : Digits): Z :=
match l with
| dsingleton d => (Z_of_Digit d)
| dcons x xs => (Z_of_Digit x) + (Z_of_Digits xs)*10
end.
\end{lstlisting}
\end{minipage}\hfill
\begin{minipage}{.4\textwidth}
\begin{lstlisting}[caption=Big endian]{bigendian}
Fixpoint Z_of_Digits (l : Digits): Z :=
match l with
| dsingleton d => (Z_of_Digit d)
| dcons x xs =>
(Z_of_Digit x)*10^(digits_len xs) +
(Z_of_Digits xs)
end.
\end{lstlisting}
\end{minipage}

The big endian function is both more computationally expensive and conceptually more complex than the little endian one. Despite that, we choose the big endian representation, because it is more natural to read numbers left-to-right, and the proofs are not much more difficult.

We also initially had two coercions \coqe{Digit} $\rightarrowtail$ \coqe{Digits} and \coqe{Digits} $\rightarrowtail \Z$ : this made writing the properties easier, however we removed it after testing the worksheets on students as they found it too confusing (in Section~\ref{sec:exp}).

\section{Binomials}
\label{sec:binom}

After a mathematical introduction, Section~\ref{sec:binom-soa} presents a
(large) state of the art about binomials.
Section~\ref{sec:binom-ct} describes numerous additional needed lemmas,
especially on sums and factorials, before going into the details of the
worksheet in Section~\ref{sec:binom-ct-ws}.

Given natural numbers~$n$ and~$k$ satisfying~$k\leq n$, the
\emph{binomial coefficient}~$n \choose k$ is the coefficient of the
monomial~$a^k b^{n-k}$ in the expansion of the binomial~$(a+b)$ to the
power~$n$, {\ie} we have
\begin{equation*}
  \forall a, b \in \R,\qquad
  (a + b)^n = \sum_{k = 0}^{n} {n \choose k} a^k b^{n-k}.
\end{equation*}
The notation~$n \choose k$ usually reads ``$n$ choose $k$''.
Indeed, the binomial coefficient also represents the number of subsets of~$k$
elements of a set of~$n$ elements.
Note that this second definition extends naturally to the {\em irregular}
case~$n<k$ with the value~0, thus making the binomial coefficient a total
function over~$\N^2$.
Note also that this combinatorial aspect is left aside here since the
formalization of subsets is itself a nontrivial issue when teaching purposes
are at stake.

Binomial coefficients appear in several fields of mathematics such as
combinatorics, probability distributions, and series expansion.
They can be generalized to real and complex numbers, by using the gamma
function instead of the factorial.
Their teaching is of great importance; it usually starts in high school, for
example for computing probabilities.

\smallskip

There are mainly two ways to define the binomial coefficients.
Either the multiplicative way using factorials or products,
\begin{equation}
  \label{eq:binom-fact}
  \forall n, k \in \N,\qquad
  {n \choose k} \EQDEF \frac{n!}{k! (n - k)!}
  \quad \mbox{when } k \leq n,\qquad
  {n \choose k} \EQDEF 0 \quad \mbox{when } n < k,
\end{equation}
or its variant where the left clause (when~$k\leq n$) is replaced with
${n\choose k}\eqdef\prod_{i=0}^{k-1}\frac{n-i}{k-i}$.
Or the additive way using sums through Pascal's rule,
\begin{equation}
  \label{eq:binom-pascal}
  \forall n, k \in \N,\qquad
  {n + 1 \choose k + 1} \EQDEF {n \choose k} + {n \choose k + 1},
\end{equation}
together with the initializations ${n \choose 0}\eqdef1$ for~$n\in\N$, and
${0 \choose k}\eqdef0$ for~$k>0$.
Of course, these definitions are equivalent, and choosing one makes the other
a lemma.
While the second one naturally defines a natural number, the first one
{\apriori} lives in the field of rational numbers.
Establishing that the division is actually exact needs a demonstration.
Note that the expression using factorials is actually valid on~$\N^2$, except
when~$(n,k)=(0,1)$, provided the use of the {\Euclidean} division and of both
conventions~$0!\eqdef1$ and $n-k\eqdef0$ when~$n<k$.

Many properties of binomial coefficients involve finite summations of natural
numbers, which are not always easy to deal with in a proof assistant.
Note also that exercises given to students often deal with general purpose
properties that may be reused in other proofs.
Then, teachers would be free to give students a truncated library file and ask
them to prove the missing results.

\subsection{State of the Art}
\label{sec:binom-soa}

As explained above, binomial coefficient are inseparable from factorials and
finite summations of natural numbers.
Several implementations of these are available through the {\Coq} community.

\smallskip

To start with, the offer provided by the {\Coq} standard library%
\footnote{\label{fn:coq-stdlib}\url{https://coq.inria.fr/doc/V8.19.1/stdlib/}.}
is not satisfactory, for three main reasons.
First, the corpus of lemmas is extremely poor: three lemmas about factorials,
five about binomial coefficients, and none about finite summations of natural
numbers.
Second, Pascal's triangle exhibits in~\eqref{eq:binom-pascal} an intrinsic
integer nature of binomials.
Thus, it is not reasonable for teaching purposes to search for assets in the
{\Reals} section of the {\Coq} standard library.
Indeed, the binomial coefficient is defined as~\coqe{C} in
\module{Reals.Binomial}, and the finite sum of endofunctions on natural numbers
as \coqe{sum_nat_f} in \module{Reals.Rfunctions}.
Moreover, binomial~\coqe{C} is defined in~$\R$ through the factorial
formula~\eqref{eq:binom-fact} and no projection on natural numbers is
provided.
And third, some lemmas could have more friendly names: \coqe{lt_0_fact} could
be renamed \coqe{fact_gt_0}, as it is a good policy to take the main notion as
prefix, and {\small \verb!pascal_step{1,2,3}!} should have more explicit
names related to their semantics.

\smallskip

\newcommand{\RSAUrl}{https://github.com/coq-contribs/rsa}
\newcommand{\BertrandUrl}{https://github.com/coq-community/bertrand}
As a consequence, several public developments in {\Coq} also provide binomial
coefficients, as they actually need them.
For instance, we may cite~{\RSA}%
\footnote{\url{\RSAUrl/}, see file
  \href{\RSAUrl/blob/master/Binomials.v}{\code{Binomial.v}}.}
for a proof of correctness of the RSA algorithm by J.~C.~Almeida and L.~Théry
in~1999,
{\Bertrand}%
\footnote{\url{\BertrandUrl/}, see files
  \href{\BertrandUrl/blob/master/theories/Summation.v}{\code{Summation.v}} and
  \href{\BertrandUrl/blob/master/theories/Binomial.v}{\code{Binomial.v}}.}
for a proof of Bertrand's postulate by L.~Théry in~2002.
Both formalize the summation $f(a)+f(a+1)+\ldots+f(a+n)$, and provide the
possibility to extract first and last terms, additivity, distributivity and
extensionality results, an induction principle, and an index shift.
{\Bertrand} adds splitting, monotony and subtractivity results, and some
properties about double summation.
Both define the binomials using Pascal's triangle~\eqref{eq:binom-pascal} and
provide basic properties such as Pascal's rule, a property using factorial, and
the binomial formula~\eqref{eq:binom-formula}.
{\Bertrand} adds positiveness, symmetry and monotony results, as well as an
upper bound when~$n$ is odd and a lower bound when it is even.

\newcommand{\FSetsUrl}{https://github.com/coq-contribs/fsets/}
We may also cite the~{\FSets}%
\footnote{\url{\FSetsUrl}, see file
  \href{{\FSetsUrl}blob/master/PowerSet.v}{\code{PowerSet.v}}.}
library for finite sets over ordered types~\cite{FL04}, in which both
definitions of the binomials are provided (using Pascal's
triangle~\eqref{eq:binom-pascal} and using factorials with the {\Euclidean}
division~\eqref{eq:binom-fact}), as well as the proof that they agree.

\smallskip

\newcommand{\CoquelicotUrl}{https://gitlab.inria.fr/coquelicot/coquelicot}
The {\Coquelicot}%
\footnote{\url{\CoquelicotUrl/}, see file
  \href{\CoquelicotUrl/blob/master/theories/Hierarchy.v}{\code{Hierarchy.v}}.}
library~\cite{BLM15} defines finite sums in the abstract setting of any abelian
monoid.
They are formalized as the summation $f(a)+f(a+1)+\ldots+f(b)$, which is zero,
the neutral element of the monoid, when $b<a$.
The support includes the extraction of first and last term, an index shift,
splitting, additivity and extensionality results.

\newcommand{\MathCompUrl}{https://github.com/math-comp/math-comp}
On the other end, the {\MathComp}%
\footnote{\url{\MathCompUrl/}, see files
  \href{\MathCompUrl/blob/master/mathcomp/ssreflect/ssrnat.v}{\code{ssrnat.v}},
  \href{\MathCompUrl/blob/master/mathcomp/ssreflect/bigop.v}{\code{bigop.v}} and
  \href{\MathCompUrl/blob/master/mathcomp/ssreflect/binomial.v}{\code{binomial.v}}.}
library~\cite{MT22} provides comprehensive support for the three features,
respectively accessible through the notations \verb|n`!| for
\coqe{factorial n}, \coqe{\sum_(m<==i<n) f} for an instantiation of
\coqe{bigop} for the addition of natural numbers, and \verb|`C(n,k)| for
\coqe{binomial n k}.

\smallskip

To sum up, none seem suitable for our teaching purposes about finite sums of
natural numbers, factorials and binomials.
The {\Coq} standard library is not competitive on these topics.
Contributions such as~{\RSA}, {\Bertrand}, or~{\FSets} would come with too many
out-of-scope lemmas.
The formalizations in {\Coquelicot}~\cite{BLM15} and~{\MathComp}~\cite{MT22}
are way too complex.
These provide desirable automation for professionals, but makes it challenging
to be used with first-years students.
We do not want to expose them to a hierarchy of abstract algebraic structures,
advanced syntax, or too general definitions.

\subsection{Provided Library about Sums, Factorials, and Binomials}
\label{sec:binom-ct}

We have developed library-like modules {\NatCompl}, {\NatSum}, {\NatFactorial}
and {\NatBinomial}, and worksheet-like modules {\WSHelper} and {\WSBinomial}.
The library modules contain definitions and the associated main properties.
The module {\NatCompl} is for complementary properties of basic operations on
natural numbers.
The module {\WSBinomial} collects proposals for exercises and their possible
solution, including statements from the library part and specific results that
only have a teaching interest.
The module {\WSHelper} aims at listing relevant results about natural numbers
for solving the exercises, and at allowing their use without qualification with
the \coqe{Nat.} prefix.
It is the only module required in {\WSBinomial}.

The library modules are only based on \module{PeanoNat}, \module{Compare\_dec}
and \module{Wf\_nat} from the {\Coq} standard library.
We provide definitions that are structurally identical to their usual
mathematical counterparts, {\ie} with increasingly indexed sums and products.
There is no addition of superficial syntax, such as implicit arguments.
There is no use of automation through tactics \coqe{auto}, \coqe{lia},
\coqe{now} or \coqe{easy}.
Thus, all proofs are completely explicit, and use plain {\Coq} style.
It is straightforward to incorporate automation in an existing proof, but the
other way around is quite cumbersome, and is provided.

We have also taken a particular care in naming the provided lemmas.
In module {\NatCompl}, the names of lemmas are prefixed by \coqe{nat_}, as if
they come from module \module{Arith.PeanoNat.Nat} from the {\Coq} standard
library.
In the other library modules, we use the following naming rules:
the main notion of the lemma is always used as prefix (here \coqe{sum_n_} and
\coqe{sum_range_} in {\NatSum}, \coqe{fact_} in {\NatFactorial} and
\coqe{binom_} in {\NatBinomial}),
results with reverse implication have suffix \coqe{_rev},
equivalence results have suffix \coqe{_equiv},
contrapositive version has suffix \coqe{_contra},
alternate version has suffix \coqe{_alt},
significant ingredients of the conclusion are used to fill the names,
and finally, we use common names when they exists.

\smallskip

The module {\NatSum} is dedicated to iterated summations on natural numbers.
Given a function~$f:\N\rightarrow\N$, among various alternatives, we choose a
definition that takes the first index~$a$ and the number of summands~$n$, and
that is structurally equal to $f(a)+f(a+1)+\ldots+f(a+n-1)$.
For convenience, we also provide the sum over a range (with both bounds
included).
\begin{lstlisting}
Fixpoint sum_n (a n : nat) (f : nat -> nat) : nat :=
  match n with O => 0 | S n' => sum_n a n' f + f (a + n') end.
Definition sum_range a b f := sum_n a (S b - a) f.
\end{lstlisting}
Thus, we have \coqe{sum_range 0 n f = sum_n 0 (S n)}.
We provide extensionality results, extraction of the first and last terms,
concatenation, splitting and extraction of any term, weak and strict monotony
results, the zero-sum property, additivity and subtractivity, left and right
distributivity of the multiplication, left and right arbitrary index
shifts, and a double summation result for the multiplication of polynomials.

\smallskip

The module {\NatFactorial} is dedicated to the support for factorials.
Given~$n\in\N$, we define \coqe{fact n} to structurally represent
$1\cdot2\cdot\ldots\cdot(n-1)\cdot n$.
The {\Coq} implementation is as follows.
\begin{lstlisting}
Fixpoint fact (n : nat) : nat := match n with O => 1 | S n' => fact n' * S n' end.
\end{lstlisting}
We provide positiveness, non-nullity and division results, weak and strict
monotony results, injectivity results, and some divisibility results.

\smallskip

The module {\NatBinomial} is dedicated to the support for binomial coefficients.
The definition follows the recurrence relation in Pascal's
triangle~\eqref{eq:binom-pascal}.
The {\Coq} implementation is as follows.
\begin{lstlisting}
Fixpoint binom (n p : nat) : nat :=
  match n, p with _, O => 1 | O, _ => 0 | S n', S p' => binom n' p' + binom n' p end.
\end{lstlisting}
Among other basic results, we provide Pascal's rule, nullity and positiveness
results, weak and strict monotony results, the committee-chair
identity~\eqref{eq:binom-cc} and its iterated version~\eqref{eq:binom-cci},
equality with the factorial formula~\eqref{eq:binom-fact}, the cancellation
identity~\eqref{eq:binom-cancel}, rising sums~\eqref{eq:binom-rising}, the
hockey stick identity~\eqref{eq:binom-hockey}, the binomial
formula~\eqref{eq:binom-formula}, Vandermonde's
identity~\eqref{eq:binom-vandermonde}, and the formulas for the sum of
binomials~\eqref{eq:binom-sum} and of their square~\eqref{eq:binom-sum-sqr}.
\begin{align}
  \label{eq:binom-cc}
  \forall n, k \in \N,\qquad& (k + 1) {n \choose k + 1} = n {n - 1 \choose k},\\
  \label{eq:binom-cci}
  \forall n, k, i \in \N,\qquad& (k \not= 0 \OR i \leq n) \IMPLIES
  (k + i)! (n - i)! {n \choose k + i} = n! k! {n - i \choose k},\\
  \label{eq:binom-cancel}
  \forall n, k, i \in \N,\qquad&
  {n \choose k + i} {k + i \choose i} = {n \choose i} {n - i \choose k},\\
  \label{eq:binom-rising}
  \forall m, n \in \N,\qquad&
  \sum_{i = 0}^n {m + i \choose m} = {m + n + 1 \choose m + 1},\\
  \label{eq:binom-hockey}
  \forall n, k \in \N,\qquad&
  \sum_{i = k}^n {i \choose k} = {n + 1 \choose k + 1},\\
  \label{eq:binom-formula}
  \forall a, b, n \in \N,\qquad&
  (a + b)^n = \sum_{i = 0}^n {n \choose i} a^i b^{n - i},\\
  \label{eq:binom-vandermonde}
  \forall m, n, p \in \N,\qquad&
  {m + n \choose p} = \sum_{i = 0}^p {m \choose i} {n \choose p - i},\\
  \label{eq:binom-sum}
  \forall n \in \N,\qquad& \sum_{k = 0}^n {n \choose k} = 2^n,\\
  \label{eq:binom-sum-sqr}
  \forall n \in \N,\qquad& \sum_{k = 0}^n {n \choose k}^2 = {2 * n \choose n}.
\end{align}

The mathematical proofs of these results do not deserve to be given here.
However, basic hints are written as comments in {\NatBinomial}, and more
detailed guidelines are provided in the worksheet {\WSBinomial}, as presented
in the next section.

\subsection{Worksheet}
\label{sec:binom-ct-ws}

Note that most general properties provided in {\NatBinomial} are commonly
considered as good exercises for the training of students.
As such, we provide a few hints and/or comments (both in French and in English)
for several lemmas on binomials, that can be transmitted to students with
statements stripped of their proofs.
The proofs in {\NatBinomial} are fully detailed, with no use of automation.
Of course, teachers are free to let students use different levels of
automation, in particular if they are sufficiently advanced in the usage of
{\Coq}.

The module {\WSHelper} provides a sufficient corpus of results from the standard
library both to reproduce the proofs of the provided library (see
Section~\ref{sec:binom-ct}) as well as to solve the proposed exercises.
Thus, it is the only module required in the worksheet {\WSBinomial}.
Teachers may tune the helper module to the exercises they extract from the
library modules.
They can also choose to hide these lemmas when students use the \coqe{Search}
command by modifying the \coqe{Add Search Blacklist} command.
We can notice that the corpus mentioned above actually behaves as a whitelist
(see the second example \coqe{binom_sum_sqr} below).

\hspace{-8pt}The worksheet {\WSBinomial} includes several exercises (lemmas) with their
possible solution (proofs).
They are equipped with comments to help students.
Comments are provided in several languages (here in English and in French), so
as not to introduce a potential additional difficulty.
Comments may be provided with several levels of mathematical details.
A script automatically generates the student version with the selected options.
It replaces the tagged solution with the command \coqe{Admitted} so that the
file is still compiling.
It is also capable of outputting \soft{HTML} using {\jsCoq}, turning the
comments into rich text using the \soft{coqdoc}~\cite{filiatre2003coqdoc} format.

\smallskip

We detail now three examples from {\WSBinomial}.
The first one, \coqe{binom_ex01a} and its variant \coqe{binom_ex01a_89},
illustrate on a binomial manipulation exercise the possibility to give students
more or less precise indications, tagged here \coqe{fr1/en1} for a precise
indication, and \coqe{fr2/en2} for more advanced students (not yet handled by
the script).
\begin{lstlisting}
(*fr L'exercice suivant provient de *) (*en The following exercise is taken from *)
(* Hyperbole Terminale - Spécialité Mathématiques, Nathan, 2020 (in French). *)

(*fr Faire la preuve du lemme suivant et remplacer "Admitted" par "Qed". *)
(*en Fill the proof of next lemma and replace "Admitted" by "Qed". *)
Lemma binom_ex01a_89 :                          Lemma binom_ex01a :
  \exists n, 5 <== n /\ binom n 5 = 17 * binom n 4.        \exists n, 5 <== n /\ binom n 5 = 17 * binom n 4.
Proof.                                          Proof.
(*fr1 Indication : essayer n = 89. *)           (*fr2 Indication : trouver une valeur pour n *)
(*en1 Hint: try n = 89. *)                      (*     en utilisant un papier et un crayon. *)
                                                (*en2 Hint: find a value for n *)
                                                (*     by using pen and paper. *)
          (* Begin solution for the teacher. *)
          exists 89. split.
          + apply sub_0_le. rewrite 5!sub_succ. apply sub_0_l.
          + apply (mul_cancel_l _ _ 5). discriminate.
            rewrite binom_mul_S_r.
            rewrite 4!sub_succ, sub_0_r, mul_assoc.
            reflexivity.
          Qed.
          (* End solution for the teacher. *)
\end{lstlisting}

For this kind of basic exercise, it could be more convenient to deal with the
factorial expression~\eqref{eq:binom-fact} with the division of real numbers
(instead of the Euclidean division).
But this would add the difficulty to deal with the injection from~$\N$
to~$\R$.

\smallskip

The second example, \coqe{binom_sum_sqr} (from the library), illustrates the
possibility to give the teacher several proofs, here with automation (on the
right) or without (on the left).
We have chosen a mild level of automation, where only trivial notions with no
consequence on the mathematical path are proved with automatic tactics (here
\coqe{lia} and \coqe{now}).
We can also imagine a version where the {\Coq} statement is written by the
student from the mathematical one~\eqref{eq:binom-sum-sqr}.
\begin{lstlisting}
(*fr Faire la preuve du lemme suivant et remplacer ``Admitted'' par ``Qed''. *)
(*en Fill the proof of next lemma and replace ``Admitted'' by ``Qed''. *)
Lemma binom_sum_sqr n : sum_range 0 n (\fun k => (binom n k) ^ 2) = binom (2 * n) n.
Proof.
(*fr Indications :
 - Transformer (2*n) en une somme.
 - Utiliser l'identité de Vandermonde.
 - Réduire l'égalité de somme à l'égalité des fonctions.
 - Utiliser la symétrie des coefficients binomiaux.
 - Terminer avec les conditions d'application des lemmes utilisés. *)
(*en Hints:
 - Transform (2*n) into a sum.
 - Use Vandermonde's identity.
 - Reduce equality on sums to equality on functions.
 - Use symmetry of binomial coefficients.
 - End with application conditions of used lemmas. *)

(* Begin solution for the teacher. *)           (* Begin other solution for the teacher. *)
(* Without automation. *)                       (* With automation. *)
replace (2 * n) with (n + n)                    replace (2 * n) with (n + n) by lia.
    by apply double_twice.                      rewrite binom_vandermonde.
rewrite binom_vandermonde.                      apply sum_range_ext. intros k Hk.
apply sum_range_ext. intros k [_ Hk].           rewrite <- binom_sym by lia.
rewrite <- binom_sym.                            now rewrite pow_2_r.
2: assumption.                                  Qed.
rewrite pow_2_r. reflexivity.                   (* End other solution for the teacher. *)
Qed.
(* End solution for the teacher. *)
\end{lstlisting}

We can note that the {\WSHelper} whitelist mentioned above can reduce here to:
\begin{lstlisting}
Export Nat (double_twice, le_add_r, lt_succ_r, pow_2_r).
\end{lstlisting}
Additional hint in the worksheet could be the output of a {\Coq} command, for
instance:
\begin{lstlisting}
(* Check binom_vandermonde:
  \forall a b n, binom (a + b) n = sum_range 0 n (\fun k => binom a k * binom b (n - k)) *)
\end{lstlisting}

\smallskip

The third example, \coqe{binom_hockey_stick} (also from the library),
illustrates the possibility to give teachers three levels of hints (from most
to least detailed) on a more elaborate property.
\begin{lstlisting}
(*fr Formule de la gouttière. *) (*en Hockey stick identity. *)
(*fr Faire la preuve du lemme suivant et remplacer "Admitted" par "Qed". *)
(*en Fill the proof of next lemma and replace "Admitted" by "Qed". *)
Lemma binom_hockey_stick n k : sum_range k n (\fun i => binom i k) = binom (S n) (S k).
Proof.
(*fr1 Indications :
 - Commencer par énoncer une version équivalente en remplaçant "sum_range k n" par
   "sum_n k p", où p représente le nombre de termes sommés, c'est-à-dire n-k+1.
   Donc "binom (S n) (S k)" devient "binom (k + p) (S k)" et il faut prouver
   "\forall p k, sum_n k p (\fun i => binom i k) = binom (k + p) (S k)".
 - Prouver cette version par induction sur p, en utilisant la formule de Pascal pour
   le cas successeur.
 - Puis faire apparaître "sum_n", utiliser le résultat prouvé précédemment, et
   discuter suivant que "S n < k" ou pas. *)
(*en1 Hints:
 - First state an equivalent version by replacing "sum_range k n" by "sum_n k p",
   where p stands for the number of added terms, ie n-k+1.
   Thus "binom (S n) (S k)" becomes "binom (k + p) (S k)" and we have to prove
   "\forall p k, sum_n k p (\fun i => binom i k) = binom (k + p) (S k)".
 - Prove this version by induction on p. Use Pascal's rule for the successor case.
 - Then, make "sum_n" visible, use the result proved previously, and discuss
   whether "S n < k" or not. *)

(*fr2 Indications :
 - Commencer par énoncer une version équivalente en remplaçant "sum_range k n" par
 "sum_n k p", où p représente le nombre de termes sommés, c'est-à-dire n-k+1,
 donc "binom (S n) (S k)" devient "binom (k + p) (S k)".
 - Prouver cette version par induction sur p (penser à la formule de Pascal).
 - Utiliser le résultat prouvé précédemment, et discuter suivant que "S n < k" ou pas. *)
(*en2 Hints:
 - First state an equivalent version by replacing "sum_range k n" by "sum_n k p",
   where p stands for the number of added terms, ie n-k+1, thus "binom (S n) (S k)"
   becomes "binom (k + p) (S k)".
 - Prove this version by induction on p (think about Pascal's rule).
 - Then, use the result proved previously, and discuss whether "S n < k" or not. *)

(*fr3 Indications :
 - Commencer par énoncer une version équivalente en remplaçant "sum_range k n" par
   "sum_n k p", où p représente le nombre de termes sommés.
 - Prouver cette version par induction sur p, puis utiliser ce résultat intermédiaire. *)
(*en3 Hints:
 - First state an equivalent version by replacing "sum_range k n" by "sum_n k p",
   where p stands for the number of added terms.
 - Prove this version by induction on p, then use this intermediary result. *)

 (* Begin solution for the teacher. *)
assert (H : \forall p k, sum_n k p (\fun i => binom i k) = binom (k + p) (S k)).
intros p. induction p.
- intros. simpl. rewrite binom_eq_0.
  reflexivity.
  rewrite add_0_r. apply lt_succ_diag_r.
- intros. simpl. rewrite IHp.
  rewrite add_succ_r. rewrite <- binom_pascal. apply add_comm.
(* *)
- unfold sum_range; rewrite H. destruct (nat_lt_le_dec (S n) k) as [H1 | H1].
  + rewrite !binom_eq_0. (* lia solves next subgoals. *)
    reflexivity.
    apply lt_trans with k. assumption. apply lt_succ_diag_r.
    rewrite (nat_sub_0_lt _ _ H1). rewrite add_0_r. apply lt_succ_diag_r.
  + f_equal. (* lia solves the next subgoal. *)
    rewrite (add_sub_assoc _ _ _ H1).
    rewrite add_comm, add_sub. reflexivity.
Qed.
(* End solution for the teacher. *)
\end{lstlisting}

Another question is the place of variables in statements.
In {\Coq}, we may write
\begin{center}
  \coqe{Lemma name : \forall n, ...}
  \hspace{1em} or \hspace{1em}
  \coqe{Lemma name n : ...}
\end{center}
Both scripts are equivalent, and have their pedagogical advantages and
disadvantages that are not discussed here.
We have made an arbitrary choice that can be modified by the teacher.

\section{Experiment}
\label{sec:exp}

In June 2024, some of us co-supervised, with Christine Paulin, two
first-year undergraduate students during two weeks.
They come from a dual degree curriculum, and had already completed a
course on formal methods using Lean and a waterproof-like library as
support. They expressed interest in an internship related to formal
methods so we took this opportunity to let them test our worksheets.

They did not have any experience using {\Coq}. We installed {\CoqIde}
on their machine, and let them finish an introductory
worksheet \cite{rousselin2024ipf} to get to know {\Coq}. In less than
two weeks, they managed to finish the two worksheets we had
created. They were very autonomous for first-year undergraduate
students.

We took into account several unsatisfactory points:
\begin{itemize}
\item The students relied on the \coqe{Check} command to interpret
what they were seeing in the goal view. This is generally fine, except
when coercions are involved. They did not
understand how statements using coercions could possibly typecheck,
and were generally very confused about them. We therefore removed the
coercions (see Section~\ref{sec:div-ex}), even though it made the
statements longer. The students preferred this, and were able to
complete the exercises.
\item Another command the students used was \coqe{Search}. While they
preferred using it rather than searching manually, they complained
that it showed too much theorems that were unrelated. Our solution to
this problem was to use the \coqe{Add Search Blacklist}
command. This removes from the search result all theorems from the
standard library. Then, we made our own module \coqe{Z}, which
contains only theorems and definitions we want the students to
use. This solves the issue, and \coqe{Search} now only shows relevant
definitions and lemmas. This is especially the case for binomials,
where the line is blurred between a (standard) library and
some of our exercises.
\end{itemize}

Another interesting point is that, for the
exercise \coqe{binom_committee_chair_alt}, they managed to complete
the proof using another longer way we had not thought about.

\section{Conclusion}
\label{sec:ccl}

We have shown a {\Coq} development about divisibility and binomials.  It
consists in additional lemmas, dedicated tactics and 49 exercises in two
worksheets.  The total is more than 250 lemmas including the 49 exercises, and
is 3,000 lines of {\Coq} long, including about 1,300 lines of specification and
comments.  Ideas of exercises either come from
textbooks~\cite{Hyp20spe,Hyp20exp} or the internet, mostly \url{bibmath.net}; a
few directly come from a worksheet of a teacher.

This library is free.  Teachers may use part or all of it, with several levels
of leeway as explained above: which exercises (from the worksheet or even the
library), which automation, which environment, which level of hints.  This pedagogical freedom is
important to adapt the contents to (i) the level of the students and (ii) the
goal of the course.  We focus on the early academic years or even a year
before, but it means a span of four years leading to more or less mature
students.  The goal may be either purely mathematics, or mainly logic or a
trade-off between both.  For example, a teacher may want to prove more advanced
lemmas with as much automation as possible while another may prevent automation
to focus on deduction rules and logical quantifiers.

A difficult point is how this work relates with the {\Coq} standard library.
It may be inconvenient to require the standard library with its bunch of
lemmas that are probably too much for a beginning student (in a \coqe{Search}
for instance).  But it is inconvenient too to not require it and have a
parallel development, to be maintained too.
Another difficult point is the level of notation and abstraction allowed.  It
is tempting to abstract mathematical objects and make the proofs once, but it
creates strange hypotheses as seen in Section~\ref{sec:div-soa}.  If a student
tries to unfold all definitions, both abstraction and notation may make the
goal impossible to read and understand.

The scope of this work is only integer properties.  But we think we have
spanned over many types of exercises, as done in mathematics.  We have studied
the ones for which we were not satisfied by the proof in order to develop
either additional lemmas or tactics to make them reachable.  The contribution
is therefore both a {\Coq} library and several {\Coq} worksheets, done with a
focus on flexibility and a pedagogical insight.

\paragraph{Perspectives}

A first perspective is of course for this material to be used more
intensively on real students. It may help sharpen some of the
exercises, by their hints, stating or expected difficulty.

A second perspective is about the script used to generate the version
for students. By now, it is designed to keep only one language and it
would be interesting to keep at least two of them. For instance, when
selecting the local language, it would be interesting to also keep
English, {\eg} for the foreign students. Moreover, as presented with
the worksheet about binomials, it will be interesting to handle the
possibility to provide several levels of indications, {\eg} with the
tags \coqe{fr1/en1}, \coqe{fr2/en2} and so on.

Another perspective is more advanced mathematics and we plan to develop a worksheet
on the Lebesgue integral.  This is at the other end of the early academic years
as possible (usually done at the end of the bachelor, three years after
\french{terminale}).  It deals with advanced concepts such as real numbers,
functions, convergence.  The mathematics usually rely on implicit axioms such
as extensionality or classical logic.  We plan to use a {\Coq} library
developed for purely research purpose~\cite{BCFMM22} and see how it can be
adapted to obtain pedagogical contents.

\section*{Acknowledgments}

We are very thankful to Gilles Delabrouille for letting us use his
paper worksheet exercises for divisibility.

We are also very thankful to Julie Delafosse and Noé Briand for testing our
worksheets during their internship.  Their insights and efficiency have
impressed us.  Finally, we are very thankful to Christine Paulin for
co-supervising the two students of the experiment.

\bibliographystyle{eptcs}
\bibliography{biblio}

\end{document}